\documentclass[
    ,final            
    ,numberedheadings 
    ,cmfonts          
  ]
  {aipproc}

\layoutstyle{6x9}

\begin{document}

\title{Astrophysical tests of \\ mirror dark matter}


\author{P.~Ciarcelluti\ }{
  address={Universit\'e de  Li\`ege, D\'epartement de  Physique B5, Sart Tilman,B-4000 LIEGE 1, Belgium}
}

E-mail: paolo.ciarcelluti@ulg.ac.be

\begin{abstract}

Mirror matter is a self-collisional dark matter candidate. If exact mirror parity is a conserved symmetry of the nature, there could exist a parallel hidden (mirror) sector of the Universe which has the same kind of particles and the same physical laws of our (visible) sector. The two sectors interact each other only via gravity, therefore mirror matter is naturally ``dark''. The most promising way to test this dark matter candidate is to look at its astrophysical signatures, as Big Bang nucleosynthesis, primordial structure formation and evolution, cosmic microwave background and large scale structure power spectra.

\end{abstract}

\maketitle


\section{Introduction}

The idea that there may exist a hidden mirror sector of particles and interactions with exactly the same properties  as our visible world 
was suggested long time ago by Lee and Yang~\cite{mirror}, and the model with exact parity symmetry interchanging corresponding fields of two sectors was proposed many years later by Foot at al.~\cite{mirror}.
The two sectors communicate with each other only via gravity\footnote{
There could be other interactions, as for example the kinetic mixing 
between O and M photons, but they are negligible for the present study.}.
A discrete symmetry $G\leftrightarrow G'$ interchanging corresponding fields of $G$ and $G'$, so called mirror parity, guarantees that two particle sectors are described by identical Lagrangians, with all coupling constants (gauge, Yukawa, Higgs) having the same pattern. 
As a consequence the two sectors should have the same microphysics.
Once the visible matter is built up by ordinary baryons, then the mirror baryons would constitute dark matter in a natural way, since they interact with mirror photons, but not interact with ordinary photons.

The phenomenology of mirror matter was studied in several papers (for an extended list see the bibliografy of ref.~\cite{okun50}), in particular the implications for Big Bang nucleosynthesis~\cite{bcv,g_mir}, primordial structure formation and cosmic microwave background~\cite{paolo,paolo1}, large scale structure of the Universe~\cite{paolo2,ignavol-lss}, microlensing events (MACHOs)~\cite{mirstar,mirMacho}.

If the mirror (M) sector exists, then the Universe along with the ordinary (O) particles should contain their mirror partners, but their densities are not the same in both sectors.
In fact, the BBN bound on the effective number of extra light neutrinos implies that the M sector has a temperature lower than the O one, that can be naturally achieved in certain inflationary models~\cite{dolgov-dnu}.
Then, two sectors have different initial conditions, they do not come into thermal equilibrium at later epoch and they evolve independently, separately conserving their entropies, and maintaining approximately constant the ratio among their temperatures.

All the differences with respect to the ordinary world can be described in terms of only two free parameters in the model, 
\begin{equation}\label{mir-param}
x \equiv \left( s' \over s \right)^{1/3} \approx {T' \over T}
~~~~~~~~~~ ; ~~~~~~~~~~
\beta \equiv \Omega'_{\rm b} / \Omega_{\rm b} ~~~~,
\end{equation}
where $T$ ($T'$), $\Omega_{b}$ ($\Omega'_{b}$), and $s$ ($s'$) are respectively the ordinary (mirror) photon temperature, cosmological baryon density, and entropy density.
The bounds on the mirror parameters are $ x < 0.7 $ and $ \beta > 1 $, the first one coming from the BBN limit and the second one from the hypothesis that a relevant fraction of dark matter is made of mirror baryons.

As far as the mirror world is cooler than the ordinary one, $x < 1$, in the mirror world all key epochs (as are baryogenesis, nucleosynthesis, recombination, etc.) proceed in somewhat different conditions than in ordinary world.
Namely, in the mirror world the relevant processes go out of equilibrium earlier than in ordinary world, which has many far going implications. 


\section{Thermodynamics of the early Universe}

Since ordinary and mirror sectors have the same microphysics, it is obvious that the neutrino decoupling temperature $T_{D\nu}$ is the same in both of them, that is $T_{D\nu}=T_{D\nu}'$. 
We use this fact together with the entropy conservation to find equations which will give the mirror photon temperature $T'$ and the ordinary and mirror thermodynamical quantities corresponding to any values of the ordinary photon temperature $T$. From them it is possible to work out the total effective number of degrees of freedom (DOF) in both sectors, which can be, as common in the literature, expressed in terms of total effective neutrino number $ N_\nu $.
The presence of the other sector indeed, leads in both 
sectors to the same effects of having more particles. 

At temperatures we are interested (below $\sim$ 10 MeV) we can in general use the following equations derived from the conservations of separated entropies:
\begin{equation}\label{eqs:1-2}
\frac{22}{21}=\frac{\frac{7}{8}q_{e}(T')+q_{\gamma}}{\frac{7}{8}q_{\nu}} \left(\frac{T'}{T_{\nu}'}\right)^3
~~~~~~ ; ~~~~~~
\frac{22}{21}=\frac{\frac{7}{8}q_{e}(T)+q_{\gamma}}{\frac{7}{8}q_{\nu}} \left(\frac{T}{T_{\nu}}\right)^3 ~~~~,
\end{equation}
and neglecting the entropy exchanges between the sectors (imposing $x$ constant):
\begin{equation}\label{eq:3}
x^3 = \frac{s'\cdot a^3}{s \cdot a^3} =
\frac{\left[ \frac{7}{8}q_e(T')+ q_{\gamma} \right] T'^3+ \frac{7}{8} q_{\nu} 
T_{\nu}'\,^3} {\left[  \frac{7}{8}q_e(T)+ q_{\gamma} \right] T^3 + 
\frac{7}{8} q_{\nu} T_{\nu}^3} ~~~~,
\end{equation}
where $s$ is the entropy density, $a$ the scale factor, $q_i$ the entropic DOF of species $i$.

At $T\simeq T_{D\nu}$ ($T'_{D\nu}$) ordinary (mirror) neutrinos decouple and soon after ordinary (mirror) electrons and positrons annihilate.

The mirror world must be colder than the ordinary one and therefore the neutrino decoupling takes place before in the mirror sector.
We can split the early Universe evolution in three phases.

\noindent 1) $T > T_{D\nu'}$:
photons and neutrinos are in thermal equilibrium in both worlds, that is 
$T_{\nu} = T \; , \; T_{\nu}' = T'$. 

\noindent 2) $T_{D\nu} < T \leq T_{D\nu'}$:
at $T \simeq T_{D\nu'}$ M neutrinos decouple and soon after M electrons and positrons annihilate, raising the M photon temperature ($T' \neq T_{\nu}'$). 
Nevertheless, O photons and neutrinos still have the same temperature 
($T = T_{\nu}$).

\noindent 3) $T \leq T_{D\nu}$:
at $T\simeq T_{D\nu}$ O neutrinos decouple and soon after O electrons and positrons annihilate, raising the O photon temperature ($T \neq T_{\nu}$).

We solved numerically equations \eqref{eqs:1-2} and \eqref{eq:3} in order to work out the total, ordinary and mirror numbers of entropic ($q$) and energetic ($g$) DOF at any temperature $T$. 
The corresponding number of neutrinos $N_{\nu}$ is found assuming that all 
particles contributing to the Universe energy density, to the exclusion of 
electrons, positrons and photons, are neutrinos; in formula that means
\begin{equation}\label{Nnu}
N_{\nu} = \frac{\bar g - g_{e^{\pm}}(T) - g_{\gamma}}{\frac{7}{8}\cdot 2} \cdot \left( \frac{T}{T_{\nu}} \right)^4 ~~~~.
\end{equation}

In table \ref{Nnu-ord} we report the asymptotic numerical results before ($ T $ = 5 MeV) and after ($ T $ = 0.005 MeV) BBN for different values of $x$. 
We stress that the standard value $N_{\nu} = 3$ is the same at any temperatures, while a distinctive feature of the mirror scenario is that the number of neutrinos raises with the temperature and with $x$.
Anyway, this effect is not a problem; on the contrary it may be useful since 
recent data fits give indications for a number of neutrinos at recent times higher than at BBN.
\begin{table}
\begin{tabular}{lrrrrrr}
\hline
   \tablehead{1}{r}{b}{T(MeV)}
  & \tablehead{1}{r}{b}{standard}
  & \tablehead{1}{r}{b}{$\mathbf{x=0.1}$}
  & \tablehead{1}{r}{b}{$\mathbf{x=0.3}$}
  & \tablehead{1}{r}{b}{$\mathbf{x=0.5}$}
  & \tablehead{1}{r}{b}{$\mathbf{x=0.7}$}   \\
\hline
5 & 3.00000 & 3.00063 & 3.04989 & 3.38430 & 4.47563\\
0.005 & 3.00000 & 3.00074 & 3.05997 & 3.46270 & 4.77751\\
\hline
\end{tabular}
\caption{Effective $ N_\nu $ in the ordinary sector.}
\label{Nnu-ord}
\end{table}

It is possible to approximate the difference between the effective numbers of neutrinos before and after the BBN process with the the follwing expression:
\begin{equation}\label{N_nu_mir_approx}
N_{\nu} (T \ll T_{ann \, e^{\pm}}) - N_{\nu} (T \gg T_{D\nu})
= x^4 \cdot \frac{1}{\frac{7}{8}\cdot 2} \left[ 10.75 - 3.36 \left( \frac{11}{4} \right)^\frac{4}{3}\right]
\simeq 1.25 \cdot x^4 ~~.
\end{equation}

The effective number of neutrinos in the mirror sector can be worked out in a similar way, and the values (much higher than the ordinary ones) are reported in table \ref{Nnu-mir}.
\begin{table}
\begin{tabular}{lrrrrr}
\hline
   \tablehead{1}{r}{b}{T(MeV)}
  & \tablehead{1}{r}{b}{$\mathbf{x=0.1}$}
  & \tablehead{1}{r}{b}{$\mathbf{x=0.3}$}
  & \tablehead{1}{r}{b}{$\mathbf{x=0.5}$}
  & \tablehead{1}{r}{b}{$\mathbf{x=0.7}$}   \\
\hline
5  & 61432 & 761.4 & 101.3 & 28.59\\
0.005 & 74011 & 917.0 & 121.4 & 33.83\\
\hline
\end{tabular}
\caption{Effective $ N_\nu $ in the mirror sector.}
\label{Nnu-mir}
\end{table}

\section{Big Bang Nucleosynthesis}

As we have seen, the presence of the mirror sector can be parametrized in terms of extra DOF number or extra neutrino families; therefore, since the physical processes involved in BBN are not affected by the mirror sector, we can use and modify a pre-existing numerical code to work out the light elements production.

The number of DOF enters the program in terms of neutrino species number; this quantity is a free parameter, but instead of using the same number during the whole BBN process, we use the variable $N_{\nu} (T, x)$ numerically computed in the previous section.
The only parameter of the mirror sector which affects ordinary BBN is $x$; the baryon ratio $\beta$ does not induce any changes on the production of ordinary nuclides, but it plays a crucial role for the mirror nuclides production.

In table \ref{tab-bbn-ord} we report the final abundances (mass fractions) of the light elements $^4He$, $D$, $^3He$ and $^7Li$ produced in the ordinary sector at the end of BBN process (at $T \sim 8\cdot 10^{-4}$ MeV) for several $x$ values and compared with the standard.
We can easily infer that for $x < 0.3$ the light element abundances do not change more than a few percent, and the difference between the standard and $x=0.1$ is of order $10^{-4}$ or less.

\begin{table}
\begin{tabular}{lrrrrrr}
\hline
  & \tablehead{1}{r}{b}{standard}
  & \tablehead{1}{r}{b}{$\mathbf{x=0.1}$}
  & \tablehead{1}{r}{b}{$\mathbf{x=0.3}$}
  & \tablehead{1}{r}{b}{$\mathbf{x=0.5}$}
  & \tablehead{1}{r}{b}{$\mathbf{x=0.7}$}   \\
\hline
$^4He$ & 0.2483 & 0.2483 & 0.2491 & 0.2538 & 0.2675\\
$D/H \: (10^{-5})$ & 2.554 & 2.555 & 2.575 & 2.709 & 3.144\\
$^3He/H \: (10^{-5})$ & 1.038 & 1.038 & 1.041 & 1.058 & 1.113\\
$^7Li/H \: (10^{-10})$ & 4.549 & 4.548 & 4.523 & 4.356 & 3.871\\
\hline
\end{tabular}
\caption{Light elements produced in the ordinary sector.}
\label{tab-bbn-ord}
\end{table}

Even mirror baryons undergo nucleosynthesis via the same physical processes than the ordinary ones, thus we can use the same numerical code also for the M nucleosynthesis.
Mirror BBN is affected also by the second mirror parameter, that is the M baryon density (introduced in terms of the ratio $\beta = \Omega'_b / \Omega_b \sim 1 \div 5$), which raises the baryon to photon ratio $\eta' = \beta x^{-3} \eta$.

The results are reported in table \ref{tab-bbn-mir}, which is the analogous of table \ref{tab-bbn-ord} but for a mirror sector with $\beta$ = 5.
We can see that BBN in the mirror sector is much more different from the standard than the ordinary sector one. This is a consequence of the high ordinary contribution to the number of total mirror DOF, which scales as $\sim x^{-4}$ (while in the ordinary sector the mirror contribution is almost insignificant, since it scales as $\sim x^4$).
Hence, in this case the M helium abundance should be much larger than that of the O helium, and for $x <0.5$ the M helium gives a dominant mass fraction of the dark matter of the Universe.
This is a very interesting feature, because it means that mirror sector can be a helium dominated world, with important consequences on star formation and evolution~\cite{mirstar}, and other related astrophysical aspects.

\begin{table}
\begin{tabular}{lrrrrrr}
\hline
  & \tablehead{1}{r}{b}{$\mathbf{x=0.1}$}
  & \tablehead{1}{r}{b}{$\mathbf{x=0.3}$}
  & \tablehead{1}{r}{b}{$\mathbf{x=0.5}$}
  & \tablehead{1}{r}{b}{$\mathbf{x=0.7}$}   \\
\hline
$^4He$ & 0.8051 & 0.6351 & 0.5035 & 0.4077\\
$D/H \: (10^{-5})$ & $1.003 \cdot 10^{-7}$ & $4.838 \cdot 10^{-4}$ & $6.587 \cdot 10^{-3}$ & $3.279 \cdot 10^{-2}$\\
$^3He/H \: (10^{-5})$ & $0.3282$ & $0.3740$ & $0.4172$ & $0.4691$\\
$^7Li/H \: (10^{-10})$ & $1.996 \cdot 10^{3}$ & $3.720 \cdot 10^{2}$ & $1.535 \cdot 10^{2}$ & $0.7962 \cdot 10^{2}$\\
\hline
\end{tabular}
\caption{Light elements produced in the mirror sector ($\beta$ = 5).}
\label{tab-bbn-mir}
\end{table}

\section{Structure formation}

The important moments for the structure formation are related to the 
matter-radiation equality (MRE) and to the matter-radiation decoupling 
(MRD) epochs. 
The MRE occurs at the redshift
\begin{equation} \label{z-eq} 
1+z_{\rm eq}= {{\Omega_m} \over {\Omega_r}} \approx 
 2.4\cdot 10^4 {{\Omega_{m}h^2} \over {1+x^4}} ~~.
\end{equation}
Therefore, for $x\ll 1$ it is not altered by the additional relativistic 
component of the M sector.
The mirror MRD temperature $T'_{\rm dec}$ 
can be calculated following the same lines as in 
the O one, obtaining $T'_{\rm dec} \approx T_{\rm dec}$, and hence 
\begin{equation} \label{z'_dec}
1+z'_{\rm dec} \simeq x^{-1} (1+z_{\rm dec}) 
\simeq 1100 \; x^{-1} ~~,
\end{equation}
so that the MRD in the M sector occurs earlier than in the O one. 

Moreover, for values 
$ x < x_{\rm eq} \simeq 0.046 \, \left( \Omega_{m} h^2 \right)^{-1}$, 
the mirror photons would decouple yet during the radiation dominated 
period. 
This critical value plays an important role in our further considerations, 
where we distinguish between two cases: $x > x_{\rm eq}$ and 
$x < x_{\rm eq}$. 
For typical values of $\Omega_{m} h^2$ we obtain $x_{\rm eq} \simeq 0.3$.

The relevant scale for gravitational instabilities is the mirror Jeans mass, 
defined as the minimum scale at which, in the matter dominated epoch, sub-horizon sized perturbations start to grow. 
In the case $x > x_{\rm eq}$ (where the mirror decoupling happens after the 
matter-radiation equality) its maximum value is reached just before 
the M decoupling, and is expressed in terms of the O one as
\begin{equation}
M_{\rm J,max}' \approx 
  \beta^{-1/2} \left( { x^4 \over {1 + x^4} } \right)^{\rm 3/2} 
  \cdot M_{\rm J,max} ~~,
\end{equation}
which, for $\beta \geq 1$ and $x < 1$, means that the Jeans mass for the M 
baryons is lower than for the O ones, 
with implications for the structure formation.
If, e.g., $ x = 0.6 $ and $ \beta = 2 $, then 
$ M_{\rm J}' \sim 0.03 \; M_{\rm J} $. 
We can also express the same quantity in terms of $ \Omega_b $, $ x $ 
and $ \beta $, in the case that all the dark matter is in the form of M baryons, as
\begin{equation} \label{mj_mir_1}
M_{\rm J}'(a_{\rm dec}') \approx 
  3.2 \cdot  10^{14} M_\odot \;
  \beta^{-1/2} ( 1 + \beta )^{-3/2} \left( x^4 \over {1+x^4} \right)^{3/2} 
  ( \Omega_{\rm b} h^2 )^{-2} ~~.
\end{equation}
For the case $ x < x_{\rm eq} $, the mirror decoupling happens before the 
matter-radiation equality.
In this case we obtain for the highest value of the Jeans mass just 
before decoupling the expression
\begin{equation}
 \label{mj_mir_2}
M_{\rm J}'(a_{\rm dec}') \approx 
  3.2 \cdot  10^{14} M_\odot \; 
  \beta^{-1/2} ( 1 + \beta )^{-3/2} 
  \left( x \over x_{\rm eq} \right)^{3/2} \left( x^4 \over {1+x^4} \right)^{3/2} 
  ( \Omega_{\rm b} h^2 )^{-2} ~~.
\end{equation}
In case $ x = x_{\rm eq} $, the expressions 
(\ref{mj_mir_1}) and (\ref{mj_mir_2}), respectively valid for 
$ x \ge x_{\rm eq} $ and $ x \le x_{\rm eq} $, are coincident, as we expect.
If we consider the differences between the highest mirror Jeans mass for 
the particular values $ x = x_{\rm eq}/2 $, $ x = x_{\rm eq} $ 
and $ x = 2 x_{\rm eq} $, we obtain the following relations:
\begin{equation}
M_{\rm J,max}'(x_{\rm eq}/2) \approx 0.005 \: M_{\rm J,max}'(x_{\rm eq})
~~~~ ; ~~~~
M_{\rm J,max}'(2x_{\rm eq}) \approx 64 \: M_{\rm J,max}'(x_{\rm eq}) ~~.
\end{equation}
Density perturbations in M baryons on scales $ M \ge M'_{\rm J,max}$, which 
enter the horizon at $z\sim z_{\rm eq}$, undergo uninterrupted linear 
growth. 
Perturbations on scales $ M \le M'_{\rm J,max}$ start instead to oscillate 
after they enter the horizon, thus delaying their growth till the 
epoch of M photon decoupling.

As occurs for perturbations in the O baryonic sector, 
also the M baryon density fluctuations should undergo the 
strong collisional Silk damping around the time of M recombination, 
so that 
the smallest perturbations that survive the dissipation will have the mass 
\begin{equation} \label{ms_m}
M'_S \sim [f(x) / 2]^3 (\beta \, \Omega_b h^2)^{-5/4} 10^{12}~ M_\odot ~~,
\end{equation}
where $f(x)=x^{5/4}$ for $x \geq x_{\rm eq}$, 
and $f(x) = (x/x_{\rm eq})^{3/2} x_{\rm eq}^{5/4}$ 
for $x \leq x_{\rm eq}$. 
For $x\sim x_{\rm eq}$ we obtain 
$ M'_S \approx 10^7 (\Omega_b h^2)^{-5} M_\odot \sim 10^{10} M_\odot $, a typical galaxy mass.

\section{Cosmic Microwave Background and Large Scale Structure}

In order to obtain quantitative predictions we computed numerically the 
evolution of scalar adiabatic perturbations in a flat Universe in which is 
present a significant fraction of mirror dark matter at the expenses of 
diminishing the cold dark matter (CDM) contribution and maintaining constant $\Omega_m$. 

\begin{figure}\label{cmblss}
  \includegraphics[height=.46\textheight]{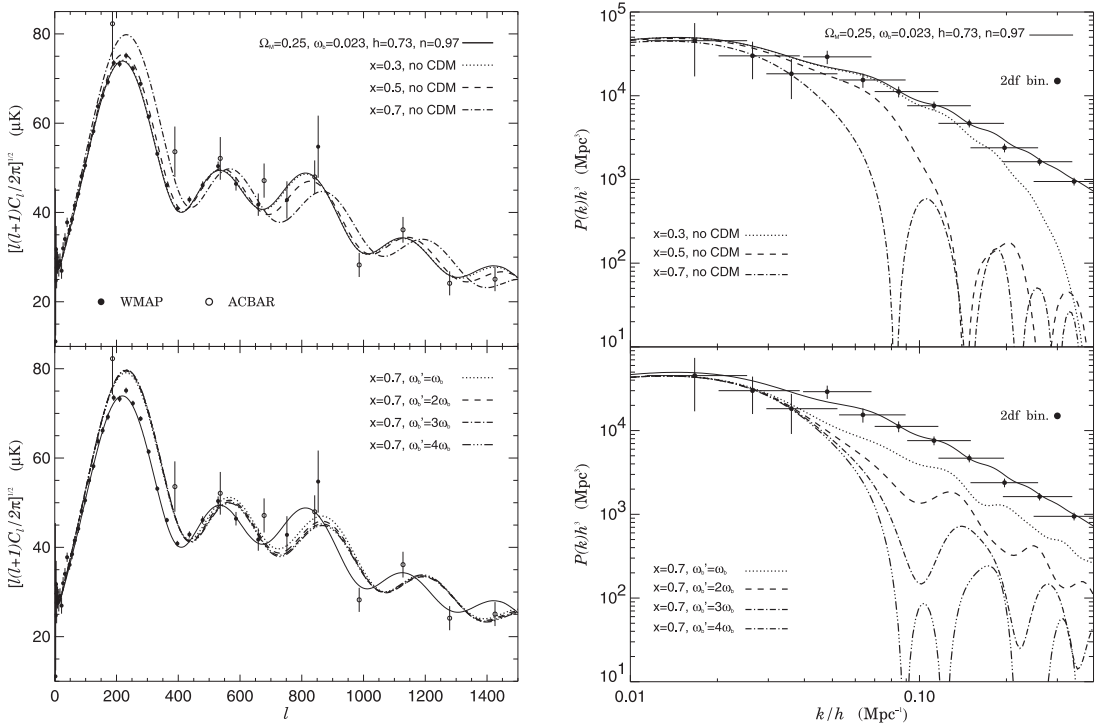}
  \caption{CMB ({\sl left}) and LSS ({\sl right}) power spectra for different values of $x$ and $\omega_{\rm b}'$, as compared with a reference standard model (solid line) and with observations.
Models where dark matter is entirely due to M baryons (no CDM) are plotted in {\sl top panels} for $x = 0.3, 0.5, 0.7$, while models with mixed CDM + M baryons ($\beta=1,2,3,4$ ; $x=0.7$) in {\sl bottom panels.} }
\end{figure}

We have chosen a ``reference cosmological model'' with the following set of parameters:
$ \omega_{b} = \Omega_{b} h^2 = 0.023, ~ 
\Omega_{ m} = 0.25, ~ 
\Omega_{\Lambda} = 0.75 , ~ 
n_{\rm s} = 0.97, ~ h = 0.73 $. 
The dependence of the CMB and LSS power spectra on the parameters 
$x$ and $\beta$ is shown in fig. \ref{cmblss}.
The predicted CMB spectrum is quite strongly dependent on the value of 
$x$, and it becomes practically indistinguishable from the CDM case for 
$x < x_{\rm eq} \approx 0.3$.
However, the effects on the CMB spectrum rather weakly depend on the fraction of mirror baryons. 
As a result of the oscillations in M baryons perturbation evolution, one observes 
oscillations in the  LSS power spectrum; their position clearly depends on $x$, while their depth depends on the mirror baryonic density.
Superimposed to oscillations one can see the cut-off in the power 
spectrum due to the aforementioned Silk damping.

In the same figure our predictions can be compared with the observational data in order to obtain some general bound on the mirror parameters space.

\begin{itemize}
\item The present LSS data are compatible with a scenario where all the 
dark matter is made of mirror baryons only if we consider enough small 
values of $ x $: 
$ x \le 0.3 \approx x_{\rm eq} $.
\item High values of $ x $, $ x > 0.6 $, can be excluded
even for a relatively small amount of mirror baryons. 
In fact, we observe 
relevant effects on LSS and CMB power spectra down to values of 
M baryon density of the order $ \Omega'_b \sim \Omega_b $. 
\item Intermediate values of $ x $, $ 0.3 < x < 0.6 $, can be 
allowed if the M baryons are a subdominant component of dark matter, 
$ \Omega_b \leq \Omega_b' \leq \Omega_{CDM} $. 
\item For small values of $ x $, $ x < 0.3 $, 
the M baryons and the CDM scenarios are indistinguishable as concerns 
the CMB and the linear LSS power spectra.
In this case, in fact, the mirror Jeans and Silk lengths, 
which mark region of the spectrum where the effects of 
mirror baryons are visible, decrease to very low values, which undergo 
non linear growth from relatively large redshift. 
\end{itemize}

Thus, with the current experimental accuracy, we can exclude only 
models with high $ x $ and high $ \Omega_b' $.

\section{Summary}

\begin{figure}\label{resume}
  \includegraphics[height=.52\textheight]{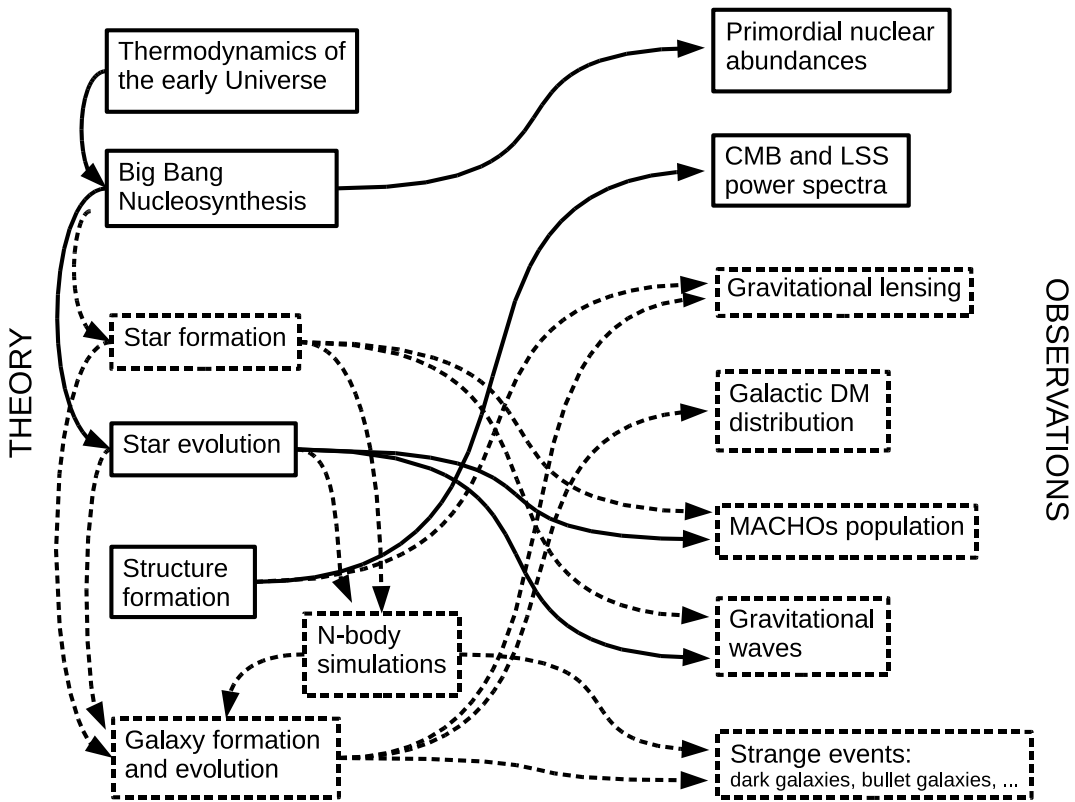}
  \caption{Current status of the astrophysical research with mirror dark 
           matter: solid lines mark what is already done, while dashed ones mark what is still to do.}
\end{figure}

Figure \ref{resume} shows the current situation of the astrophysical research in presence of mirror dark matter.
We have already investigated the early Universe (thermodynamics and Big Bang nucleosynthesis), and the process of structure formation in linear regime, that permit to obtain predictions, respectively, on the primordial elements abundances, and on the observed cosmic microwave background and large scale structure power spectra.
In addition, we have studied the evolution of mirror dark stars, which, together with the mirror star formation, are necessary ingredients for the study and the numerical simulations of non linear structure formation, and of the formation and evolution of galaxies; furthermore, in future studies they will provide predictions on the observed abundances of MACHOs and on the gravitational waves background.
Ultimately we will be able to obtain theoretical estimates to be compared with observations of gravitational lensing, galactic dark matter distribution, and strange astrophysical events still unexplained (as for example dark galaxies, bullet galaxy, ...).

Concluding, the astrophysical tests so far used show that mirror matter can be a viable candidate for dark matter, but we still need to complete the entire picture of the Mirror Universe.


\begin{theacknowledgments}
This work was supported by the Belgian Science Policy Office Inter
University Attraction Pole VI/11 ``Fundamental Interactions''.
\end{theacknowledgments}


\begin{thebibliography}{9}

\bibitem{mirror}
T.~D.~Lee, and C.~N.~Yang,  \emph{Phys. Rev.}, \textbf{104}, 254--258 (1956); \\
R.~Foot, H.~Lew, and R.R.~Volkas, \emph{Phys. Lett. B}, \textbf{272}, 67--70 (1991).
\bibitem{okun50}
L.~B.~Okun, hep-ph/0606202.
\bibitem{bcv}
Z.~Berezhiani, D.~Comelli and F.~L.~Villante, \emph{Phys. Lett. B}, \textbf{503}, 362--375 (2001) [hep-ph/0008105].
\bibitem{g_mir}
P.~Ciarcelluti, A.~Lepidi, arXiv:0809.0677.
\bibitem{paolo}
P.~Ciarcelluti, \emph{PhD thesis}, [astro-ph/0312607]; \\
Z.~Berezhiani, P.~Ciarcelluti, D.~Comelli, and F.~Villante, \emph{Int. J. Mod. Phys. D}, \textbf{14}, 107-120 (2005) [astro-ph/0312605]; \\
P.~Ciarcelluti, \emph{Frascati Phys. Ser.}, \textbf{555}, 225--228 (2004) [astro-ph/0409629].
\bibitem{paolo1}
P.~Ciarcelluti, \emph{Int. J. Mod. Phys. D}, \textbf{14}, 187--222 (2005) [astro-ph/0409630].
\bibitem{paolo2}
P.~Ciarcelluti, \emph{Int. J. Mod. Phys. D}, \textbf{14}, 223--256 (2005) [astro-ph/0409633].
\bibitem{ignavol-lss}
A.Yu.~Ignatiev, and R.R.~Volkas, \emph{Phys. Rev. D}, \textbf{68}, 023518 (2003) [hep-ph/0304260].
\bibitem{mirstar}
Z.~Berezhiani, P.~Ciarcelluti, S.~Cassisi, A.~Pietrinferni, \emph{Astropart. Phys.} \textbf{24}, 495-510 (2006) [astro-ph/0507153].
\bibitem{mirMacho}
S.~Blinnikov, astro-ph/9801015; \\
R.~Foot, \emph{Phys. Lett. B}, \textbf{452}, 83--86 (1999) [astro-ph/9902065]; \\
R.~N.~Mohapatra, V.~Teplitz, \emph{Phys. Lett. B}, \textbf{462}, 302--309 (1999) [astro-ph/9902085].
\bibitem{dolgov-dnu}
Z.G.~Berezhiani, A.D.~Dolgov, and R.N.~Mohapatra, \emph{Phys. Lett. B}, \textbf{375}, 26--36 (1996) [hep-ph/9511221].

\end{thebibliography}
\end{document}